\documentclass{article}
\usepackage{xcolor}
\usepackage{INTERSPEECH2018,amsmath,epsfig}
\usepackage{url}

\title{NISP: A Multi-lingual Multi-accent Dataset for Speaker Profiling}
\name{  
\\
 Shareef Babu Kalluri$^1$,  Deepu Vijayasenan$^1$,  Sriram Ganapathy$^2$, \\ Ragesh Rajan M$^1$, Prashant Krishnan$^2$}
\address{$^1$National Institute of Technology Karnataka, Surathkal, India, \\
  $^2$Learning and Extraction of Acoustic Patterns (LEAP) lab, Indian Institute of Science, Bangalore.}
\email{\{shareefbabu1,deepu.senan,sriram.iisc,mrageshrajan,gillyprash29\}@gmail.com}
\begin{document}
\ninept
\maketitle
\begin{abstract}
Many commercial and forensic applications of speech demand the extraction of information about the speaker characteristics, which falls into the broad category of speaker profiling. The speaker characteristics needed for profiling include physical traits of the speaker like height, age and gender of the speaker along with native language of the speaker. Many of the datasets available have only partial information for speaker profiling. In this paper, we attempt to overcome this limitation by developing a new dataset which has speech data from five different Indian languages along with English.  The metadata information for speaker profiling applications like linguistic information, regional information and physical characteristics of a speaker are also collected.  We call this dataset as \textit{NITK-IISc Multilingual Multi-accent Speaker Profiling} (NISP) dataset. The description of dataset, potential applications and baseline results for speaker profiling on this dataset are provided in this paper. 
\end{abstract}
\section{Introduction}
\label{sec:intro}
\textcolor{black}%
{In the recent years, speech is emerging as a reliable biometric  for various commercial and surveillance applications. Speech contains the speaker identity  information along with textual information, 
geographical information (region from where the individual belongs to) in the form of accent, age (child / teenager / adult), gender (male / female), social information, and also the emotional state of the person (angry, happy, sad, anxious etc.) \cite{rabiner1978digital}.  Extraction of  speaker related meta information is known as speaker profiling. This metadata can be used in commercial applications like voice agents and dialog systems, to deliver content targeted to the user~\cite{schuller}.   
Also, in forensic scenarios, speaker profiling could provide clues about the caller.  . Such applications have resulted in increased interest in  area of speaker profiling~\cite{poorjam2014multitask} and it makes creation of datasets in this domain very essential.  Building effective speaker profiling 
systems require large amount of good quality speech data along with metadata such as gender, age, physical characteristics, accent. }

 \textcolor{black}%
 {Existing speech corpora has limited information about speaker metadata.  Most of them 
 have either physical characteristics or accent information, but often not about both. 
 For example, the most common dataset TIMIT~\cite{timit}  has only age height and 
 gender information about the speakers. There is no information about other physical 
 parameters or about the accents.   The popular Speaker Recognition Evaluation (SRE) 
 challenge datasets~\cite{SRE,SRE08,SRE10} have in addition the information about 
 smoking habits and native country. They don't have linguistic information.   Other datasets such as 2010 Interspeech
 Paralinguistic Challenge(ComParE) dataset~\cite{schuller},  Fisher English Corpus \cite{cieri2004fisher},
 SpeechDat II dataset \cite{speechdat}  provide only the gender and age group
 information of the speaker. The CMU Kids~\cite{cmukids} dataset just provides the grade 
 in which the kids are studying. None of 
these datasets provide any details about physical parameters beyond height and age. 
The only exception to this is the Copycat corpus \cite{lehman2016estimation} that has
 details of height, weight and age, but the speakers are limited to children. 
Similarly there are also data sets that provide the only the accent information
of the speakers such as  Accents of British Isles (ABI-1) corpus~\cite{d2004accents} and the CSLU-Foreign Accent English (FAE) \cite{FAE} datasets.  In this context, there is a 
need for dataset with richer metadata for speaker profiling systems.}

 Another limitation of current datasets is that most of the available datasets are monolingual (English). On the other hand, multi-lingual data available (for example, the Babel dataset \cite{harper2013babel}) do not have detailed speaker profiling information. In order to build a speaker profiling system robust to multiple languages and accents, we require a dataset that also has all the required speaker profiling metadata information along with the speech data. 

In this paper, we attempt to overcome some of the limitations of the available datasets by collecting multilingual, multi-accent dataset from five Indian states. This dataset is  called \textit{NITK-IISc Multilingual Multi-accent Speaker Profiling} (NISP) dataset. \footnote{This dataset is publicly made available in the following address,   \textit{https://github.com/iiscleap/NISP-Dataset}. This dataset is freely available for academic and  research purposes only  with standard license agreements.} 
We describe the details of the dataset  in this paper along with baseline systems for speaker profiling. 

The rest of the paper is organized as follows.  Sec.~\ref{sec:design of database} describes the design and description of the dataset. Sec.~\ref{sec:stats} provides details about the statistics of the dataset. Sec.~\ref{sec:applications} provides the  list of potential applications where NISP data can be useful.  Sec.~\ref{sec:baseline} contains the discussion on the baseline experimental results on physical parameter estimation.  This is followed by a discussion and summary in Sec. \ref{sec:concl}.
\section{Design of Database }
\label{sec:design of database}
\subsection{Metadata}
The NISP dataset creation involved collecting the speech and metadata from Indian speakers belonging to five Indian languages. The entire data collection took place over the course of a year. 
The speakers who participated in contributing speech data for this database consisted of students, academic staff and faculty members  of 
different educational institutions across southern India. An informed consent is obtained from the speakers 
to use the data for academic and research activities. 
The linguistic, regional and physical traits are collected from each speaker  along with the speech data. The metadata information collected in this dataset are the following,
 \begin{enumerate}
\item  Native language (L1) of the speaker and whether the speaker can read text from L1. 
\item {Language used  in the schooling years}.
\item {Second language (L2)} - Most commonly spoken language other than L1.
\item {Regional Information}: The geographic location of the native place (or the place where the subject has lived dominantly). 
\item Current place of residence.
\item {Physical Characteristics Information:} Age, gender of the speaker and body build parameters like height, shoulder size, and weight.
The age of the speaker was noted in years and the height is measured in centimetres.
The shoulder size of the speaker is measured at the widest point of shoulders between acromion bone with the individual's arms at their side in centimetres.
And the weight of a speaker is measured in kilograms using  standard digital weighing machine.
\end{enumerate}

\subsection{Speech data}
The audio recordings were collected in a quiet environment  like a normal class room or seminar hall in each of the 
educational institutions.  All necessary precautions are taken care to avoid ambient noise, and reverberations. Also any fans or air conditioners were switched off during the data collection process. The speech data was collected using a high quality microphone (with Scarlett solo studio, CM25 a large diaphragm condenser microphone  ). The data was sampled at $44.1$ kHz with a bit-rate $16$ bits per sample. In order to avoid any channel variations across recordings, all the speech samples were collected using the same microphone device. 

The text data used in the reading task for the speakers were presented in the L1 language as well as in English in two 
different sessions.  The text provided to speakers were taken from the daily news articles as unique sentences without any contextual continuity  from one sentence to another in both native language and English texts. This setting was  made 
to avoid any prosodic continuity in the reading  task. Separately, a  continuous short story section was given to the 
speakers in both the L1 language and English language to have contextual continuity effects in the reading task. 
Along with these sentences, we had also used five common sentences for every speaker. This includes two TIMIT \textit{sa1} and \textit{sa2} 
sentences and three general news article sentences in English language (to perform speaker profiling in text dependent manner). Similarly two common sentences were also made in the native language text. Overall, each subject provided  $20$-$25$ unique   sentences in L1 and English, $20$-$25$ 
contextual sentences in L1 and English,  $5$ common sentences for English, and $2$ sentences from L1. Each speaker was instructed to read aloud in a clear voice with a  close talking microphone. 
\begin{table}
	\centering
	\caption{Distribution of native languages', and the number of male and female speakers in the NISP dataset  }
	\label{tab:NISP_distribution}
	\vspace{-8pt}
	\renewcommand{\arraystretch}{1.1}
	
	\begin{tabular}{|c|c|c|c|c|} 
		\noalign{\smallskip} \hline  
		Sl.No. & Native Language & Male & Female & Total \\ \hline
		1. & Hindi &  76 & 27 & 103 \\ \hline
		2. & Kannada & 33 & 27 & 60 \\ \hline
		3. & Malayalam & 35 & 25 & 60 \\ \hline
		4. & Telugu &  35 & 22 & 57 \\ \hline
		5. & Tamil  & 40 & 25 & 65 \\ \hline
		\multicolumn{2}{|c|} {Total Speakers }  & 219 & 126 & 345 \\ \hline
	\end{tabular}
\end{table}  

\subsection{Recording Protocol}
The audio recording setup is made by using a publicly available software, namely ``\textit{Speech Recorder}'' 
\footnote{This software is available in this address,   \textit{https://www.bas.uni-muenchen.de/forschung/Bas/software/speechrecorder/}} and with \textit{Focusrite Scarlett solo studio} audio recording device by connecting it to a laptop.  This audio recorder device has gain controller to adjust the gain and amplitude of the speech signal while recording. The software enables a graphical user interface (GUI) to display each sentence at a time on the screen of the speaker and it is monitored and controlled by a controller on  another display. The participant is asked to read out the text aloud which is displayed on the monitor in a comfortable sitting posture.  The controller also verified  the content, which is being read, in order to avoid any reading errors made by the speaker.  

The technical specifications and statistics of collected datasets are detailed in the following section.
\begin{figure}[h]
	\centering{
		\includegraphics[width=\linewidth]{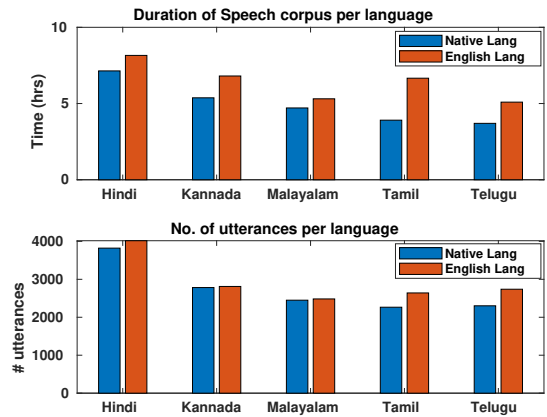}}
		\vspace{-20pt}
	\caption{Number of utterances and speech duration of each language (both native language and English speech data) in  the  NISP dataset}
	\label{fig:dur_utt}
\end{figure}
 
\section{Dataset Statistics}
\label{sec:stats}
The NISP dataset has $345$ speakers, which includes $219$ male and $126$ female speakers. The dataset has five native Indian languages (namely Hindi, Kannada, Malayalam, Tamil and Telugu) as well as Indian accented English. 
Each speaker provided around $4$-$5$ minutes of speech data in each language. 
The distribution of speakers across the different native  languages as well as gender wise distribution is shown in Table~\ref{tab:NISP_distribution}. The total number of utterances in this dataset are $28,268$, out of which $17,844$ are male speaker utterances, and $10,424$ are female speaker utterances. The total number of native language utterances are $13577$ and there are $14691$ English utterances in the dataset. This dataset has a total of $24.83$ hours of native language speech data and $32.03$ hours of English speech data.
\begin{figure}[h]
	\centering{
		\includegraphics[width=\linewidth]{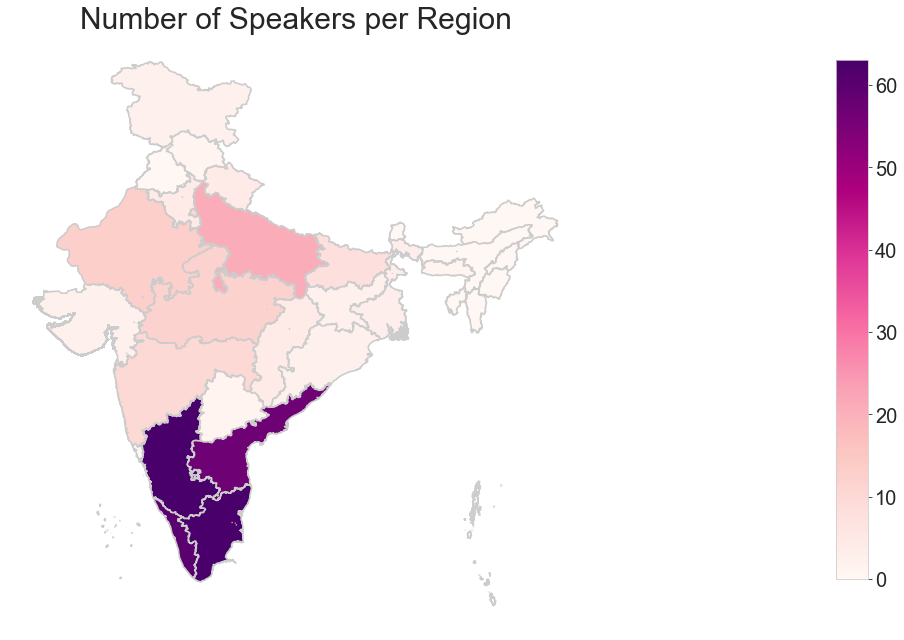}}
	\caption{ Native geographic region of the speakers in the NISP dataset.}
	\label{fig:heatmap}
\end{figure}
	
The total duration of speech in hours and total number of utterances  corresponding to each native language along with English speech are shown in Fig \ref{fig:dur_utt}.   The gender wise statistics of each physical parameters are given in Table \ref{tab:NISP_stsats}. 
The total number of speakers from each region per accent is shown in Fig \ref{fig:heatmap}.
\begin{table}
	\caption{Gender wise statistics of each physical parameter in the NISP dataset} 	
	\label{tab:NISP_stsats}
	\vspace{-20pt}
	\begin{center}
		\begin{tabular}{lrrrr}
			\noalign{\smallskip}\hline\noalign{\smallskip}
			\textbf{Physical} & \textbf{Min}  & \textbf{Max}   &  \textbf{Mean}    &   \textbf{Standard} \\
			\textbf{Characteristic} & & & & \textbf{Deviation} \\			 
			\noalign{\smallskip}\hline\noalign{\smallskip}\multicolumn{5}{c} {Male Speakers}\\			
			\noalign{\smallskip}\hline\noalign{\smallskip}
			Height (\emph{cm}) & 151.0 & 191.0 & 171.6 & 6.7 \\
			Shoulder width (\emph{cm}) &  32.0 &  55.0 &  44.7 & 3.2 \\
			Weight (\emph{kg}) &  43.4 & 116.5 &  69.4 & 11.9 \\
			Age (\emph{y}) &  18.0 &  47.5 &  24.4 & 5.6 \\

			\noalign{\smallskip}\hline\noalign{\smallskip}\multicolumn{5}{c} {Female Speakers}\\			
			\noalign{\smallskip}\hline\noalign{\smallskip}
			Height (\emph{cm}) & 143.0 & 180.0 & 158.9 &  6.8 \\
			Shoulder width (\emph{cm}) & 30.0 & 53.0 & 39.7 &  3.4 \\
			Weight (\emph{kg}) & 34.1 & 86.2 & 56.5 & 10.5 \\
			Age (\emph{y}) & 18.3 & 46.5 & 25.1 &  6.1 \\
			
			\noalign{\smallskip}\hline\noalign{\smallskip}	
			\multicolumn{5}{c} {Male and Female Speakers }\\	
			\noalign{\smallskip}\hline\noalign{\smallskip}
			Height (\emph{cm}) & 143.0 & 191.0 & 166.9 & 9.1 \\
			Shoulder  width (\emph{cm}) & 30.0 & 55.0 & 42.9 & 4.0 \\
			Weight (\emph{kg}) & 34.1 & 116.5 & 64.7 & 13.0 \\
			Age (\emph{y}) & 18.0 & 47.5 & 24.7 & 5.8 \\
			\noalign{\smallskip}\hline\noalign{\smallskip}							
		\end{tabular}
	\end{center}
\end{table}

\section{Potential Applications } 
\label{sec:applications}
The NISP dataset provides a wide range of various applications depending on the task requirement. This dataset provides the insight to explore more about the multilingual setting of speaker profiling applications in text dependent or independent fashion, accent/language identification experiments, speaker recognition as well as multilingual speech recognition experiments. 

\subsection{Physical Parameter Estimation}
As most of the available speaker profiling databases  are specific to one language (English), this developed NISP dataset has speech data with multiple native  languages of India  (Hindi, Kannada, Malayalam,Tamil and Telugu ) along with English speech recordings from each native speaker. 
\subsection{Accent \& Language Identification}
Identifying the accent and L1 of the speaker is an important cue in the voice forensic applications as well as in smart speaker and dialog systems. 
The NISP dataset enables research to explore accent related effects on speech. This database  allows both L1 identification from L2 as well as language identification based on the $5$ L1 languages. Note that many of L1 languages are from  geographically connected regions of the country and therefore we hypothesize language identification will itself be challenging in this setting. 
\subsection{Speaker Recognition}
The large scale speaker recognition datasets  \cite{Nagrani17,Chung18b} etc.,) are monolingual (English). Many of these datasets are currently used to build large neural network based embedding extractors. The NISP dataset, while being much smaller in scale, can be used to fine-tune the large neural network models with more multi-accent and multi-lingual variabilities. We hypothesize that this can improve the robustness of speaker recognition systems. In addition, multilingual speaker verification with mismatched languages in enrollment and test data can be useful for bench-marking speaker verification systems.
\subsection{Speech Recognition}
This dataset has potentially rich text information in both English and all the native  languages (Hindi, Kannada, Malayalam, Tamil and Telugu). All these transcription, after manual verification, are recorded in UTF-8 format. The dataset also enables accented speech recognition research along with multi-lingual ASR experiments. 
\begin{figure}[t!]
	\centering{
	\hspace{-0.5 cm}	\includegraphics[width=\linewidth,height=5cm]{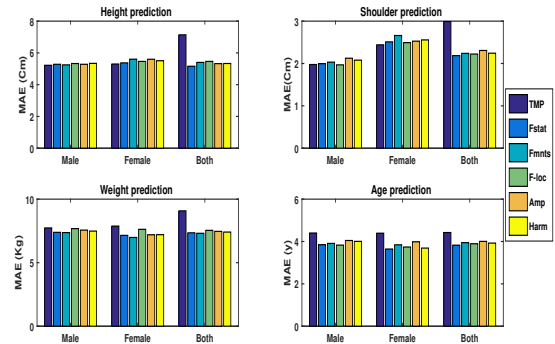}
			}
	\caption{Gender wise MAE of each feature (Fstat,Formants (Fmnts), frequency locations (F-loc), Amplitude (Amp) and harmonic features (amplitude $+$ frequency locations -- Harm )) compared with Training data Mean Predictor (TMP) of the NISP dataset}
	\label{fig:MAE_NISP}
	\vspace{-10pt}
\end{figure}
\section{Baseline Experiments and Results }
\label{sec:baseline}
For the evaluation purposes, the dataset is divided into train and test splits without overlapping any speakers.  The training split has 210 speakers with 17161 utterances, which comprises of 134 male speakers with 10911 utterances and 76 female speakers with 6250 utterances. For test split, there are 135 speakers with 11107 utterances, which includes 85 male speakers with 6933 utterances and 50 female speakers with 4174 utterances.   The statistics of train and test splits of the dataset are given in Table~\ref{tab:NISP_stsats_trts}. The standard  error metrics Mean Absolute Error (MAE) and Root Mean Square Error (RMSE) are used to measure the errors from the actual and predicted targets. 

We estimate the physical parameters like height, age, shoulder size and weight using the NISP dataset.  We perform the physical parameter estimation task using three different features namely, mel filter bank features, formants and harmonics. More details about the feature extraction setup is given in \cite{kalluri2020automatic}. We computed the first order statistics (Fstat) from the $40$ Mel filter bank features using a $256$ component diagonal covariance Gaussian Mixture Model Universal Background Model (GMM-UBM). The GMM was trained using $20$ Mel Frequency Cepstral Coefficients (MFCC)  and its deltas and double deltas together constitutes $60$ dimensional  features. The formant and fundamental frequency features are extracted using wide band spectral components with $18^{th}$ order all pole model. The percentiles ($5$,$25$,$50$,$75$ and $95$) are computed for the extracted features over the entire utterance. Also the harmonic features including both frequency locations (F-loc) and amplitude features  (Amp.) are extracted using the narrow band spectral components using $80^{th}$ order all pole model.  The same set of percentiles are computed for the harmonic features over the entire utterance. These computed statistics from each individual feature are given to linear Support Vector Regression (SVR) model to predict each physical parameter.  

The MAE of each individual  feature is shown in Fig \ref{fig:MAE_NISP}. This is compared with the default approach - the Training data Mean Predictor (predicting the target physical parameter using the mean of training data of each parameter). We performed the simple average of predicted targets of these individual regression outputs of these features to improve the performance of the final predicted targets.   

\begin{table}
	\caption{Statistics of Train and Test splits  of each physical parameter in the NISP dataset} 
	\label{tab:NISP_stsats_trts}
	\vspace{-20pt}
	\begin{center}
		\begin{tabular}{lrrrr}
			\noalign{\smallskip}\hline\noalign{\smallskip}
			\textbf{Physical} & \textbf{Min}  & \textbf{Max}   &  \textbf{Mean}    &   \textbf{Standard} \\
			\textbf{Characteristic} & & & & \textbf{Deviation} \\			 
			\noalign{\smallskip}\hline\noalign{\smallskip}\multicolumn{5}{c} {Train Speakers}\\			
			\noalign{\smallskip}\hline\noalign{\smallskip}
			Height (\emph{cm}) & 143 & 191 & 167.1 & 9.5 \\
			Shoulder width (\emph{cm}) & 32 & 55 & 42.9 & 4.2\\
			Weight (\emph{kg}) &  36.9 & 116.5 & 65.4 & 14.0 \\
			Age (\emph{y}) &  18 & 47.5 & 24.8 & 6.0\\

			\noalign{\smallskip}\hline\noalign{\smallskip}\multicolumn{5}{c} {Test Speakers}\\			
			\noalign{\smallskip}\hline\noalign{\smallskip}
			Height (\emph{cm}) & 146.5 & 182.5 & 166.7 & 8.5 \\
			Shoulder width (\emph{cm}) & 30.0 & 53.0 & 42.9 & 3.7 \\
			Weight (\emph{kg}) & 34.1 & 93.8 & 63.5 & 11.3 \\
			Age (\emph{y}) & 18.3 & 43.6 & 24.4 & 5.5 \\
			
			\noalign{\smallskip}\hline\noalign{\smallskip}							
		\end{tabular}
	\end{center}
\end{table}
The three different Support Vector Regression outputs of first order statistics, formants and the harmonic features (both frequency and amplitude features) were combined (Comb--3).  These results are tabulated in comparison with default predictor in Table \ref{tab:NISP_comb_res}.
This simple average of  predicted targets of these features has improved the predicted error metrics over the individual error metrics. 
The MAE and RMSE of both speakers (male and female speakers) improved relatively by about $22-29$\%  in body build parameter estimation (height, shoulder width and weight) tasks. Similarly, in age estimation, we observe a relative improvement of $14$\% improvement in MAE. There is a relative improvement over the TMP with three feature combination (Comb--3) in all the physical parameters except in RMSE of female speakers' shoulder size and male speakers' age.                                                                                                                                                                                                                                                                                                                                                                                                                                                                                                                                                                                                                                                                                                                                                                                                                                                                                                                                                                                                                                                                                                                                                                                                                                                                                                                                                                                                                                                                                                                                                                   
\begin{table}
	
	\caption{Comparison of three  feature combinations with default predictor --  Comb -3 (Fstats $+$ formant $+$ harmonic features (amplitude $+$ frequency locations)) }
\label{tab:NISP_comb_res}
	\renewcommand{\arraystretch}{1.0}
	\centering
	\begin{tabular}{r c c c c c c}
		\noalign{\smallskip}\hline\noalign{\smallskip}\noalign{\smallskip}
		\multicolumn{7}{c} {Height (cm) Estimation}\\
		\noalign{\smallskip}\hline\noalign{\smallskip}
		& \multicolumn{2}{c}{Male}  &\multicolumn{2}{c}{Female} & \multicolumn{2}{c}{All} \\
		\noalign{\smallskip}\hline\noalign{\smallskip}\noalign{\smallskip} 
		& MAE & RMSE & MAE & RMSE & MAE & RMSE \\ \noalign{\smallskip}\hline\noalign{\smallskip}
		TMP & 5.22 & 6.17 & 5.30 & 6.93 & 7.14 & 8.47 \\
		Comb--3 & \textbf{5.16} & \textbf{6.13} & \textbf{5.30} & \textbf{6.70} & \textbf{5.11} & \textbf{6.15} \\
		\noalign{\smallskip}\hline\noalign{\smallskip}
		\multicolumn{7}{c} {Shoulder (cm) Estimation}  \\
\noalign{\smallskip}\hline\noalign{\smallskip} \noalign{\smallskip}
TMP & 1.98 & 2.58 & \textbf{2.44} & \textbf{3.52} & 2.99 & 3.73\\
Comb--3 & \textbf{1.93} & \textbf{2.48} & 2.47 & 3.55 & \textbf{2.11} & \textbf{2.85} \\
				\noalign{\smallskip}\hline\noalign{\smallskip}
		\multicolumn{7}{c} {Weight(kg) Estimation}  \\
\noalign{\smallskip}\hline\noalign{\smallskip}
		TMP & 7.74 & 9.57 & 7.88 & 9.76 & 9.08 & 11.35 \\
		Comb--3 & \textbf{7.06} & \textbf{8.79} & \textbf{6.84} & \textbf{8.61} & \textbf{7.06} & \textbf{8.78} \\
		\noalign{\smallskip}\hline\noalign{\smallskip} \noalign{\smallskip}
		\multicolumn{7}{c} {Age(y) Estimation}  \\
		\noalign{\smallskip}\hline\noalign{\smallskip} \noalign{\smallskip}
TMP & 4.40 & \textbf{5.60} & 4.39 & 5.57 & 4.42 & 5.54 \\
Comb--3 & \textbf{3.80} & 5.63 & \textbf{3.55} & \textbf{4.99} & \textbf{3.76} & \textbf{5.48}\\

		\noalign{\smallskip}\hline\noalign{\smallskip}
	\end{tabular}
		\hspace{-30pt}
\end{table} 
	\hspace{-30pt}

\section{Conclusions}\label{sec:concl}
A multilingual speaker profiling dataset is presented in this paper where the data was recorded in five different Indian native languages (Hindi, Kannada, Malayalam, Tamil, and Telugu) along with English language. This dataset has the linguistic information, regional information and physical characteristics of a speaker which are all useful in commercial and forensic applications of speaker profiling. This dataset has $345$ ($219$ males and $126$ females) speakers and contains $28,268$ ($17,844$ from male speaker, and $10,424$ from female speaker) utterances. Overall, this dataset has $56.86$ hours of speech data in which $24.83$ hours of data came from native languages of the speaker and $32.03$ hours of English data. For speaker profiling tasks on this dataset, the baseline results with the combination of three features (Fstats, formants and harmonics) performs better in MAE and RMSE measures when compared to the training mean predictor.

\section{Acknowledgments}
This work was partially funded by Science and Engineering Research Board (SERB) under grant no: EMR/2016/007934. 
Authors would like to acknowledge support from institutions namely, National Institute of Technology Karnataka (NITK) Surathkal, Indian Institute of Science (IISc)  Bangalore, Sree Vidyanikethan Engineering College, Tirupathi, Andhra Pradesh, KSR  College of Engineering,  Tiruchengode, Tamilnadu, and College of Engineering Thalassery, Kerala. We also acknowledge support from staff and students from these institutions for smooth conduction of data collection.

\bibliographystyle{IEEEbib}
\bibliography{refs}

\end{document}